\documentclass[fleqn,review,3p]{elsarticle}
\usepackage{multirow}
\usepackage[algoruled, linesnumbered]{algorithm2e}
\usepackage{cite}
\usepackage{graphicx}
\usepackage{psfrag}
\usepackage{epsfig}
\usepackage{subfigure}
\usepackage{url}
\usepackage{amsmath}
\usepackage{array}
\usepackage{amssymb}
\usepackage{graphicx}
\usepackage{psfrag}
\usepackage{epsfig}
\usepackage{subfigure}
\usepackage{url}
\usepackage{stfloats}
\usepackage{array}
\usepackage{slashbox}
\usepackage{soul,color}
\usepackage{hyperref}
\hypersetup{
   bookmarks=true,         
   unicode=false,          
   pdftoolbar=true,        
   pdfmenubar=true,        
   pdffitwindow=false,     
   pdfstartview={FitH},    
   pdftitle={My title},    
   pdfauthor={Author},     
   pdfsubject={Subject},   
   pdfcreator={Creator},   
   pdfproducer={Producer}, 
   pdfkeywords={keyword1} {key2} {key3}, 
   pdfnewwindow=true,      
   colorlinks=true,       
   linkcolor=blue,          
   citecolor=blue,        
   filecolor=magenta,      
   urlcolor=blue           
}

\journal{}

\usepackage{amsmath}

\begin{document}

\begin{frontmatter}

\title{Lung sound classification using local binary pattern}

\author[iitkgp]{Nandini Sengupta\corref{cor1}}
\ead{nandinisengupta@iitkgp.ac.in}
\author[uef]{Md Sahidullah}
\ead{sahid@cs.uef.fi}
\author[iitkgp]{Goutam Saha}
\ead{gsaha@ece.iitkgp.ernet.in}
\address[iitkgp]{Department of Electronics and Electrical Communication Engineering, Indian Institute of Technology, Kharagpur, India, Kharagpur-721 302.}
\address[uef]{Speech and Image Processing Unit, School of Computing, University of Eastern Finland, Joensuu, Finland, Finland-80101.}

\cortext[cor1]{Corresponding Author}

\begin{abstract}
Lung sounds contain vital information about pulmonary pathology. In this paper, we use short-term spectral characteristics of lung sounds to recognize associated diseases. Motivated by the success of auditory perception based techniques in speech signal classification, we represent time-frequency information of lung sounds using mel-scale warped spectral coefficients, called here as mel-frequency spectral coefficients (MFSCs). Next, we employ local binary pattern analysis (LBP) to capture texture information of the MFSCs, and the feature vectors are subsequently derived using histogram representation. The proposed features are used with three well-known classifiers in this field: k-nearest neighbor (kNN), artificial neural network (ANN), and support vector machine (SVM). Also, the performance was tested with multiple SVM kernels. We conduct extensive performance evaluation experiments using two databases which include normal and adventitious sounds. Results show that the proposed features with SVM and also with kNN classifier outperform commonly used wavelet-based features as well as our previously investigated mel-frequency cepstral coefficients (MFCCs) based statistical features, specifically in abnormal sound detection. Proposed features also yield better results than morphological features and energy features computed from rational dilation wavelet coefficients. The Bhattacharyya kernel performs considerably better than other kernels. Further, we optimize the configuration of the proposed feature extraction algorithm. Finally, we have applied mRMR (minimum-redundancy maximum-relevancy) based feature selection method to remove redundancy in the feature vector which makes the proposed method computationally more efficient without any degradation in the performance. The overall performance gain is up to $24.5\%$ as compared to the standard wavelet feature based system.

\end{abstract}


\end{frontmatter}


\section{Introduction}\label{Section:Introduction}

Lung disease is the third largest cause of death in the world. According to the World Health Organization (WHO) in 2015, 3.17 million people died due to \emph{chronic obstructive pulmonary diseases} (COPDs), 3.19 million people died because of \emph{lower respiratory infections} and 1.69 million people died due to \emph{trachea bronchus, lung cancer}\footnote{\url{http://www.who.int/mediacentre/factsheets/fs310/en/}}.
Various physiological and external reasons cause structural changes and abnormalities in the respiratory system. Those abnormalities can be detected through several techniques such as X-ray~\citep{wipf1999diagnosing}, CT-scan,  arterial-blood gas analysis~\citep{gould1991lung}, spirometry, peak-flow meter, stethoscope etc.

Though these processes are good to get information about lung status, they suffer from several practical problems. First of all, these techniques are not widely accessible. Other than availability, the techniques themselves are not without limitations. Arterial blood gas analysis is invasive and expensive; CT-scan or X-ray radiation is harmful to the body, spirogram (graph generated by spirometry) depends on the cooperation and effort of subjects~\citep{schulz2007spirometry}, and peak expiratory flow rate also depends on the subject's cooperation. The \emph{auscultation} method through stethoscope has several limitations too due to several issues like (a) a lack of experience of the physician leading to his inability to identify the lung sound abnormalities and impairments, and (b) low sensitivity of the human ear to the lower frequency band present in lung sound~\citep{Kandaswamy2004523}. The sound has also been captured from mouth using microphone~\citep{lei2014content}. Now the low cost, non-invasive and easily-available stethoscope-based technique with computerized assistance removes the manual interference and makes it automatic by employing machine learning techniques.

In~\citep{sengupta2016lung}, there is a discussion about common types of lung sounds - \emph{normal}, \emph{continuous adventitious sounds} (CASs) (e.g., wheeze) and \emph{discontinuous adventitious sounds} (DASs) (e.g., crackle) and their spectral characteristics~{\citep{NEJMra1302901}}. In short, normal sounds are related to the lung sounds of a healthy subject. The normal sound becomes adventitious when it is superimposed with other sounds due to different types of abnormalities or diseases in the respiratory system. Out of several types of lung diseases, basic two types of diseases are \emph{airway related diseases} and \emph{lung tissue related diseases}. Airway related diseases cause obstruction or blockage in the airways. A common symptom of obstructive diseases (e.g., asthma, COPD, etc.) is wheeze sound. Other than asthma and COPD, infections such as \emph{croup}, \emph{whooping cough}, \emph{laryngitis}, \emph{acute tracheobronchitis}, \emph{tracheal stenosis}, \emph{laryngeal stenosis} and \emph{airway compression} are also associated with wheezing sound~\citep{meslier1995wheezes}. Besides, in case of lung tissue related diseases, scarring or inflammation of the tissue makes the lungs unable to expand fully. Thus, it becomes hard for the lungs to take in oxygen and release carbon dioxide. \emph{Interstitial lung disease} (ILD) is synonymous with diffuse parenchymal lung disease, for a large group of lung diseases affecting the tissue and space around the air sacs or alveoli of the lungs, which cause progressive scarring of lung tissue through inflammation and fibrosis~\citep{zibrak2014interstitial}. One of the physical signs in ILD is crackles~\citep{piirila1995crackles,charleston2011assessment}. In this work, we focus on characterizing normal and abnormal sounds. We conduct experiments with two databases where the abnormal sounds of the first database having wheeze and crackle type lung sounds. On the other hand, the abnormal lung sounds are recorded from ILD patients in the second database.

\par
As a typical pattern recognition problem, the automatic lung sound recognition system involves three different steps: \emph{pre-processing}, \emph{feature extraction} and \emph{classification}. In the pre-processing step, the signal recorded using stethoscope is processed for heart sound effect reduction~\citep{sengupta2016lung}. In the feature extraction step, lung sound signal is represented by its features defining a set of measured values with distinctive information. Finally, in the classification step, the extracted features are categorized into normal and different diseased classes. Various feature extraction and classification techniques have been used to investigate the distinctive characteristics of lung sound. A detailed study of features and classification methods have been reported in~\citep{Palaniappan2013129}. Different features are extracted after analyzing the sound characteristics with different analysis methods described in~\citep{reichert2008analysis}. Both time and frequency domain features are used in lung sound detection. According to the previous works in this research area, the most popular features are based on wavelets~\citep{serbes2011feature,pittner1999feature,lu2008integrated,lin2006wheeze}.
They captures useful time-frequency information from the non-stationary lung sounds. Along with the wavelet-based features, spectral features are also widely used in lung sound characterization~\citep{Abbas2010Auto,riella2009method,Xie2012MSPCA,jin2011adventitious,rietveld1999classification,waitman2000representation,munakata1991spectral,jin2011adventitious}. \emph{Auto-regressive} (AR) coefficients, related to the linear prediction (LP) analysis have been used for this purpose~\citep{sankur1994comparison,alsmadi2008design,Chang2010141,martinez2006computerized,charleston2011assessment,cohen1984analysis,charleston2011assessment}.
Cepstral features such as \emph{mel-frequency cepstral coefficients} (MFCCs)~\citep{sahidullah2012design} and \emph{perceptual linear prediction cepstral coefficients} (PLPCCs) are motivated by auditory perception and are employed for this task~\citep{Bahoura2009824,orjuela2014artificial,chien2007wheeze,bahoura2004respiratory,mayorga2010acoustics,lei2014content,sengupta2016lung,sengupta2015optimization}. The auditory perception based features, which are expected to capture important distinctive characteristics of different lung sounds similar to an expert physician, show considerable lung sound recognition accuracy~\citep{sengupta2016lung,Bahoura2009824,lei2014content}. Besides, diseases like ILD which actually represents a group of lung diseases has also been detected by analyzing different features captured from lung sounds~\citep{charleston2011assessment,palaniappan2014pulmonary}.

\par
In our recent work, we have proposed a mean-based representation of perceptual features for lung sound classification that captures statistical information of cepstral coefficients~\citep{sengupta2016lung}. In this work, we investigate a new method by texture-based representation of perceptual features. In image processing, texture-based approaches are widely used for representation and classification of natural images~\citep{tuceryan1993texture}. The texture was also introduced by Ren$\mathrm{\acute{e}}$ Laennec, the inventor of the stethoscope, as a set of descriptors for the characterization of lung sounds~\citep{laennec1838treatise}. Influenced by the idea of Lanec, lacunarity feature is utilized in~\citep{hadjileontiadis2009texture}. Lacunarity is one kind of texture feature and it yields good recognition accuracy~\citep{hadjileontiadis2009texture}. Other than lacunarity, different morphological features like skewness, sample entropy are also employed for lung sound detection~\citep{bhattacharyya2015novel,jin2014new}. But, these texture features are generally extracted directly from time domain lung sound signal. Now, we can assume that lung sound may also have some texture characteristics in the frequency domain.

In this work, we investigate a new feature for lung sound characterization which captures texture-related information from its auditory perceptual representation. The features are extracted from the short-term spectral coefficients using \emph{local binary pattern} (LBP)~\citep{ojala2002multiresolution,huang2011local,pietikainen2011local,shan2009facial,shan2005robust,liao2009dominant,heikkila2009description}. LBP is widely used in image processing literature for texture analysis. In speech signal processing, LBP was used to capture texture information from speech signal for spoofing detection~\citep{alegre2013new,wu2015spoofing}, speaker recognition~\citep{roy2012fast}. We introduce this first for lung sound characterization. In this paper, LBP is used to take the texture information out from the auditory scale warped lung sound spectrum. We utilize kNN classifier for this purpose. We also use the extracted features with conventional \emph{artificial neural-network} (ANN) as it is popularly used for lung sound characterization~\citep{Palaniappan2013129,Kandaswamy2004523}. We further employ \emph{support vector machine} (SVM) as a classifier as the LBP-based feature yields better performance with SVM classifier~\citep{chapelle1999support,wong2006application}. Other than the simple linear inner product kernel, we also have used \emph{Bhattacharyya}~\citep{kondor2003kernel} and \emph{intersection} kernel~\citep{maji2013efficient} which seem to be more appropriate for the newly investigated feature. We have found that the proposed feature performs better than the  wavelet-based, morphological and our previously proposed mean-based MFCC features.
\par

The rest of the paper is organized as follows.
At first, in Section~\ref{Feature for lung sound analysis}, extraction of suitable feature for lung sound analysis and formulation of the proposed feature are discussed. After that, the basic mathematical background of used classifiers are noted down in Section~\ref{Classifiers for Lung Sound Classification}. Experimental setups are described in Section~\ref{Experimental setup}. Results are discussed in Section~\ref{result}. In Section~\ref{conclusion}, we provide a summary of the findings along with the limitations of the current work and potential future directions.

\section{Local binary pattern (LBP) based feature for lung sound analysis}\label{Feature for lung sound analysis}

Lung sound signals are not stationary as the volume of the lung is varying in nature due to change in air pressure and its amount~\citep{reichert2008analysis}. The non-stationarity is severe in case of abnormal subjects~\citep{Kandaswamy2004523}. Hence, the time-frequency representation of lung sounds can capture useful information related to lung status. In our previous study~\citep{sengupta2016lung}, by performing F-ratio analysis on short-term power spectrum, we have shown that the power spectrum contains discriminative information about different lung sounds. Subsequently, we had proposed to use the statistical information of short-term features for lung sound characterization. In this current work, we have investigated a LBP-based approach to process the short-term time-frequency representation of lung sounds. LBP is a non-parametric descriptor, which summarizes the local structure of an image by comparing each pixel with its neighboring pixels. The computational simplicity is one of the main strengths of LBP~\citep{huang2011local}, and it is a powerful approach to describe local structures. Here, we first describe the calculation of LBP which is followed by the feature vector formulation from LBP.

\subsection{Mel-frequency spectral coefficients (MFSCs)}

Proposed feature is formulated by considering several factors discussed earlier. Assuming the non-stationary nature of lung sound, we have used short-time processing~\citep{sengupta2016lung}. Secondly, considering the success of the auditory perceptual processing in lung sound classification purpose~\citep{Bahoura2009824,sengupta2016lung}, mel-scale warping is applied on short-time power spectrum. Now the mel-scale warped short-term representation can be viewed as a 2D image whose dimensions are represented by time (or number of frames) and frequency bins (or number of filters). Finally, we can use this image for extracting relevant texture information using LBP. Note that we do not need to apply \emph{discrete cosine transform} (DCT) to extract cepstral features from mel-scale warped power spectrum. Rather, we compute LBP directly on the spectrum, i.e., \emph{mel-frequency spectral coefficients} (MFSCs). This MFSC is same as mel-frequency log-energy or MFLE as discussed in~\citep{sahidullah2012design}.

In order to compute MFSCs, short-term power spectrum is calculated from lung sound frame, $y(n)$, using $N_{FFT}$-point FFT as:

\begin{equation}
|Y(k)|^2=\bigg|\sum_{n=0}^{L_w-1} y(n)\cdot e ^{\big(\frac{-j2\pi nk}{N_{FFT}}\big)}\bigg|^2, \label{eqn2}
\end{equation}

where, $1\leq k \leq \frac{N_{FFT}}{2}+1$ and $L_W$ is the window length. Next, filterbank with non-linearly spaced triangular mel-filters is imposed on the spectrum. Mel-scale is defined as,

\begin{equation}
f_\mathrm{mel}{(f)}=2595\log_{10}\big(1+\frac{f}{700}\big)\label{eqn1}
\end{equation}

where $f$ is the original frequency in Hz.

The outputs $\{ e(i) \} _{i=1} ^{Q}$ of the mel-filters can be calculated by a weighted summation between respective filter response $\psi_i(k)$ and the energy spectrum $|Y(k)|^{2}$ as:

\begin{equation}
e(i)=\sum_{k=1}^{\frac{N_{FFT}}{2}+1} |Y(k)|^2\cdot \psi_i(k) \label{eqn3}
\end{equation}

Here, $Q$ is the number of filters in filter bank.
Finally, $\log$ filterbank energies $\{\log[e(i)] \} _{i=1} ^{Q}$ is computed and we get the mel-frequency spectral coefficients.

\subsection{Computation of LBP}

The original LBP operator labels the image pixels with decimal numbers~\citep{ojala2002multiresolution}. Each pixel is compared with its neighboring pixels primarily with ${(3\times3)}$ neighborhood by subtracting the center pixel value~\citep{alegre2013new}. The resulting strictly negative values are encoded with $0$ and the other with $1$. At first, it was only for ${(3\times3)}$ neighborhood which is small enough for large images. But then it was generalized with the neighbor of different sizes. Formally, let a pixel at ${(x_c,y_c)}$, then the LBP can be expressed in decimal form as,
\begin{equation}
LBP_{P,R}(x_c,y_c)=\displaystyle\sum_{p=0}^{P-1} S{(i_p-i_c)}2^{p},
\end{equation}

where $i_p$ and $i_c$ are intensities (gray values) of $p$-th pixel and center pixel. $(P,R)$ represents a neighbourhood of $P$ sampling points on a circle of radius $R$. $S{(x)}$ function can be defined as,
\begin{equation}
S{(x)}=\begin{cases}
    1 \quad \text{if}\,\, x\geq 0 \\
    0 \quad \text{if}\,\, x< 0
      \end{cases}.
    \end{equation}

The operator $LBP_{P,R}$ produces $2^{P}$ different binary patterns formed by $P$ pixels in the neighborhood.

%
%
%
%


\subsection{Feature vector formulation using LBP}
\begin{figure}
\centerline{\includegraphics[width=8cm]{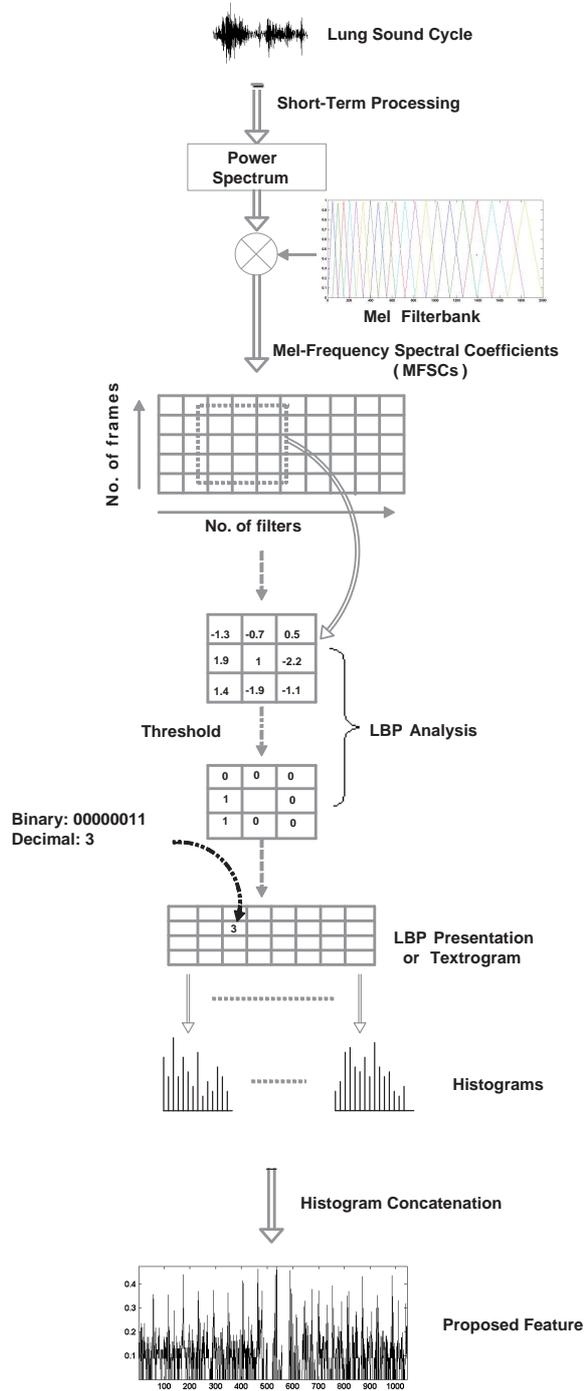}}
\caption{An illustration of the processing steps for proposed feature extraction method: from the lung cycle to histogram.}\label{figure 4}
\end{figure}

The steps of feature extraction are summarized in Fig.~\ref{figure 4}. Here, MFSCs are computed from short-time power spectrum of $T$ frames using $Q$ filters. The frame-wise concatenated MFSCs are considered as image and LBP analysis is employed on it. The standard LBP operator~\citep{ojala2002multiresolution} is a non-parametric, $3 \times 3$ kernel which assigns a binary code to each pixel in an image according to the comparison of its intensity value to that of its eight surrounding pixels. A binary value of ‘1’ is assigned when the intensity of neighboring pixels (i.e., MFSC feature) is higher, whereas a value of ‘0’ is assigned when neighboring pixels are of lower or equal intensity. Each pixel is thus assigned one of $2^{8} = 256$ binary patterns. In this work, we reduce the number of possible patterns according to the standard Uniform LBP approach reported in~\citep{ojala2002multiresolution}. Uniform LBPs are the subset of $58$ patterns which incorporate at most two bitwise transitions from $0\rightarrow1$ or $1\rightarrow0$ when the bit pattern is traversed in circular fashion.

\begin{figure}[t]
\centerline{\includegraphics[width=16cm]{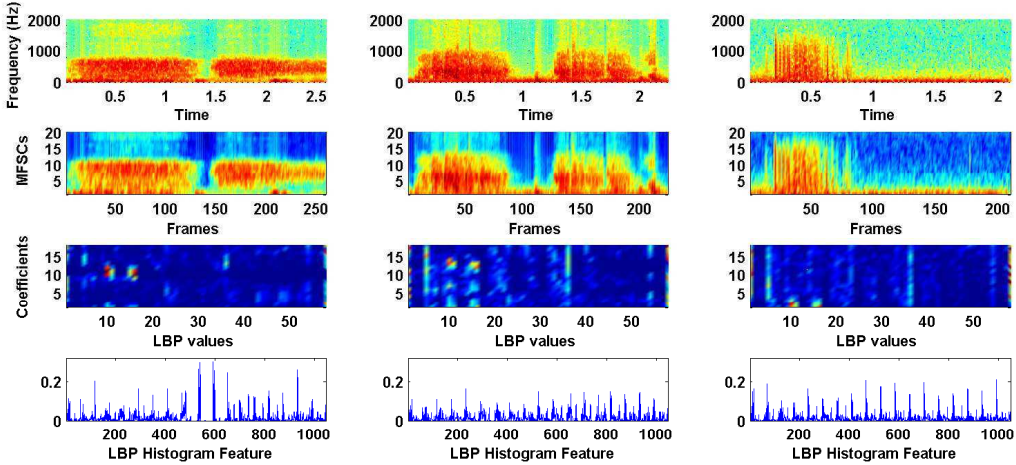}}
 \caption{Feature formulation steps showing spectrogram (top row), MFSCs and LBP (second from below) for normal (left), wheeze (center) and crackle (right). The last row shows the corresponding feature vector.}\label{Ftr_formulation}
\end{figure}

Figure~\ref{Ftr_formulation} gives the pictorial representation of the main steps of proposed feature formulation. The left column is the representation for normal sound; the middle column is of wheeze sound, and the column on the right represents the steps for crackle sound. The first row (upper one) illustrates the spectrograms for normal, wheeze and crackle sound which makes it clear that crackle has higher frequency components than normal or wheeze and normal lung sound has a narrower frequency range than other two. In the middle row, we have plotted MFSCs. Its x-axis represents the frames and y-axis represents the coefficients. This MFSC domain is a linear mapping from spectrogram. The later is more compact than the spectrogram plot as the number of filters is less than the number of frequency points in power spectrum computation. Then, the third row is a mapping from MFSCs to LBP domain. Finally, we have plotted LBP histogram features for each of the classes.



%

As per the description in~\citep{ojala2002multiresolution,alegre2013new}, most patterns are naturally uniform and according to the different pieces of evidence, many image recognition applications lead to better performance by using only uniform patterns than the full set of uniform and non-uniform patterns. Thus pixels corresponding to any of the 198 non-uniform patterns are ignored in our work too. LBPs are computed for each pixel in the mel-frequency spectral coefficients. Therefore, a new matrix referred to as a \emph{textrogram} which is of the reduced dynamic range is generated. The LBP-based feature is created by concatenating histograms configured from the pixel values across each filter in the textrogram. However, the textrograms corresponding to the filters in beginning and end are discarded as they do not have all the neighbors. The LBP feature vector is formed by normalization of the histograms individually.

In our work, we have generally used $20$ filters in mel scale. Therefore, unless otherwise specified, the feature dimension is $(20-2) \times 58 = 1044$.

\begin{figure}[t]
\centerline{\includegraphics[width=14cm]{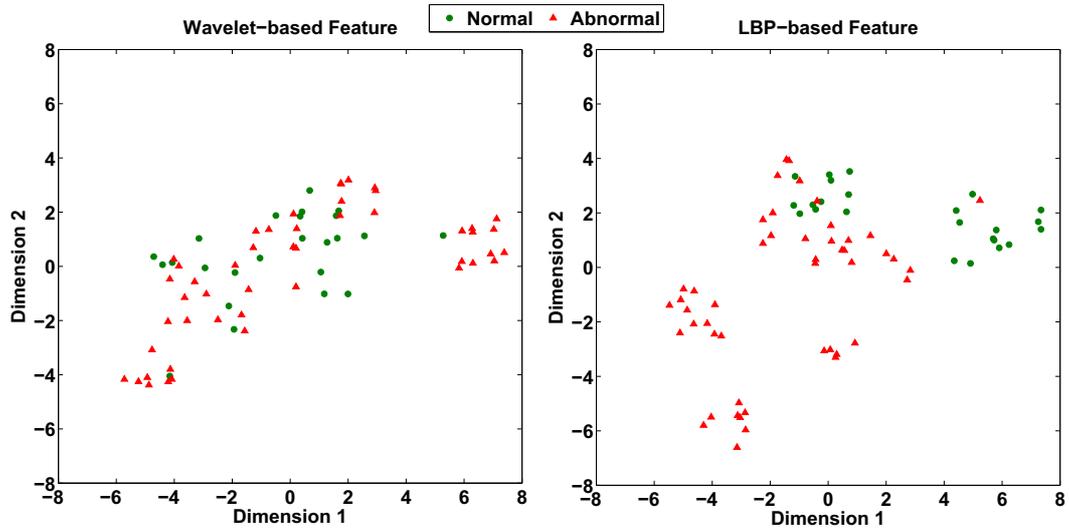}}
 \caption{Scatter plot showing the normal (\textit{green circle}) and abnormal sounds (\textit{red triangle}) for all the lung sound cycles in Database 1 (Details in Section~\ref{Label_Database}).}\label{tSNE}
\end{figure}

Figure~\ref{tSNE} shows an illustration of scatter plot of feature vectors using wavelet-based and LBP-based approach for normal and abnormal lung sounds. Two-dimensional data representation is produced using t-distributed stochastic neighbor embedding (t-SNE)~\citep{maaten2008visualizing} algorithm applied to feature vector of each lung sound cycle. Two classes of sounds are visually more separable when LBP-based features are used. Also, we have done MANOVA analysis to figure out the separability between the classes. The Wilk’s lambda values of the databases are shown in Table~\ref{manova}. Null hypothesis is closer to rejection when Wilk’s lambda is close to zero. Thus, it can be inferred that classes are more separable when LBP based feature is used.

%

\begin{table}[t]
\centering
\caption{MANOVA analysis: Wilk’s lambda value (p-value $<$ $0.05$) of corresponding databases are represented.}
\label{manova}
\begin{tabular}{|c|c|c|}
\hline
{Feature} & Database 1    & Database 2   \\ \hline
Wavelet                  & 0.4923      & 0.1652    \\ \hline
LBP                      & 0.1507      & 0.0501     \\ \hline
\end{tabular}
\end{table}

\section{Classifiers for lung sound classification}\label{Classifiers for Lung Sound Classification}

\subsection{k-nearest neighbour(kNN)}
\begin{figure}[h]
\centering
\hspace{-0.8cm} \includegraphics[width=8cm,height=4cm]{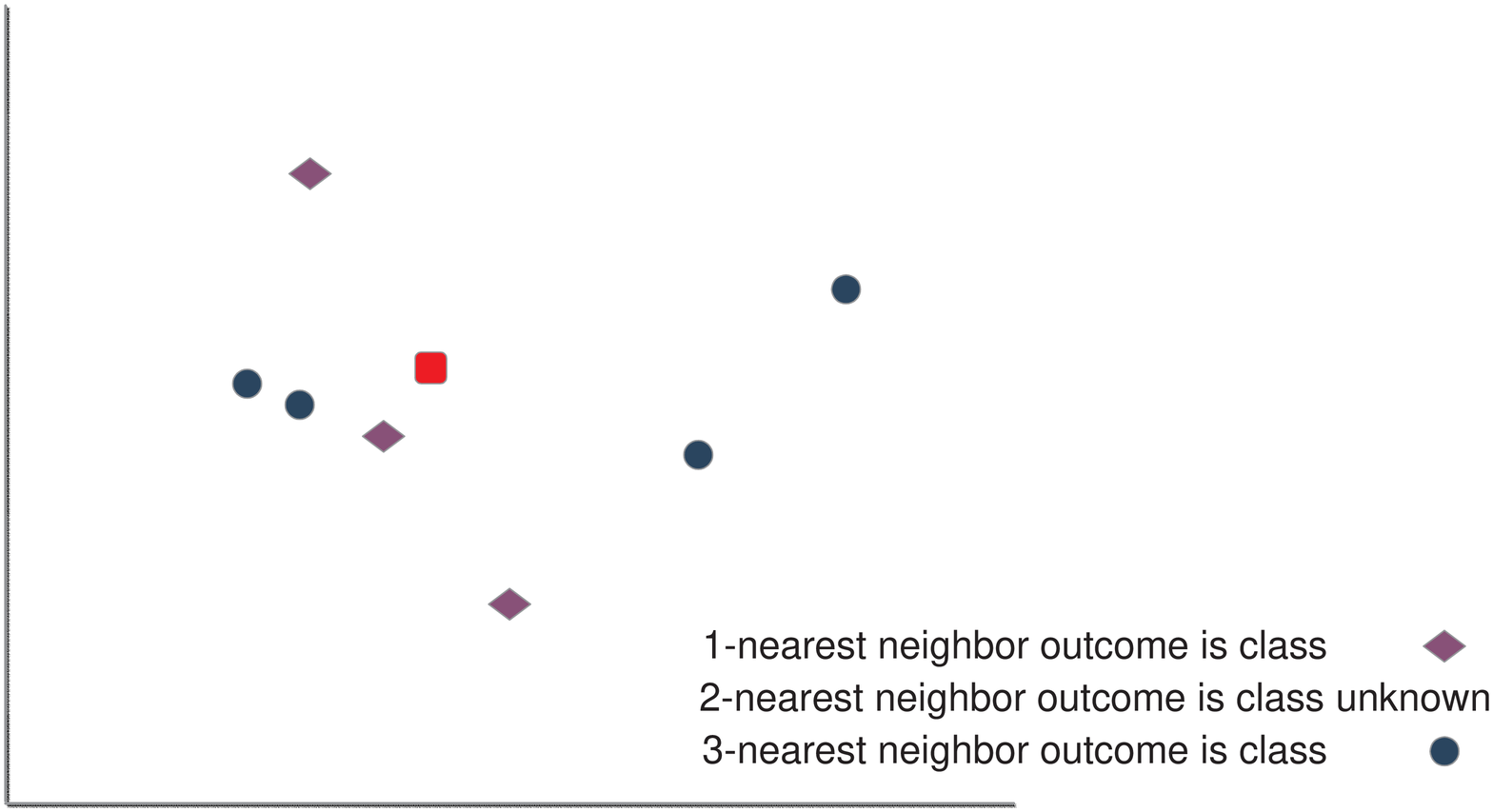}\\
 \caption{A representative diagram of kNN classifier.}\label{KNN}
\end{figure}

kth nearest neighboring (kNN) is a nonparametric classifier. To demonstrate a k-nearest neighbor analysis, let's consider the task of classifying an unknown object among some known objects. The training stage of the kNN algorithm comprises of storing the feature vectors and class labels of the training samples. In the classification stage, the same features are used as before for the the unknown test sample. Distances from the unknown input vector to all stored vectors are computed, and k closest samples are selected. The unknown  sample is predicted to belong to the particular class that is the most numerous within the set. The Euclidean distance is typically used to measure the distance or similarity between instances. However, other distance functions can also be employed for this purpose. Figure~\ref{KNN} depicts a simple representative figure of a nearest neighbor classifier.

\subsection{Artificial neural network (ANN)}

\begin{figure}
\centering
\hspace{-0.8cm} \includegraphics[width=6cm,height=5cm]{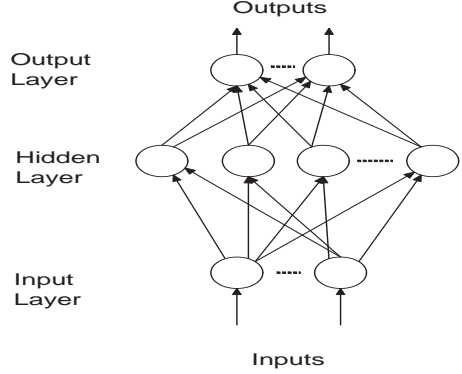}\\
 \caption{A representative diagram of ANN classifier.}\label{ANN1}
\end{figure}

The ANN used here is a \emph{multi-layer perceptron} (MLP) which is a number of neurons (nodes), arranged together in layers in feed-forward manner~\citep{SimonHaykin}. It consists of three layers -- input, hidden and an output layer. Inputs pass through first two layers and finally emerge from the output layer. Figure \ref{ANN1} represents an example of an MLP network characterized by one input layer, one hidden layer, and one output layer.

Each node, in the hidden layer, receives the output of each node from the input layer through a connection of weight and produced response is forwarded to the output layer. Each node of hidden layer performs a weighted sum which is transferred by a nonlinear function and its results proceeds to the output layer.
\begin{equation}
     h_j=f_h(\displaystyle\sum_{i=0}^{d} w_{j,i}x_i)
\end{equation}

     where, $h_j$ is the produced response of $j$-th node of hidden layer, $f_h$ is the non-linear function at the hidden layer node, $d$ is number of nodes in input layer, $w_{j,i}$ is the weight connecting $i$-th input node and $j$-th hidden layer node, $x_i$ is the i-th input feature of input feature vector, $x_0=1$ is the bias term.

Same ways, response of hidden layer passes through another non-linear function after multiplied by the weights of the output layer.
     \begin{equation}
     o_k=f_o(\displaystyle\sum_{j=0}^{m}w_{k,j}(h_j))
     \end{equation}

    where, $o_k$ is the produced response of $k$-th node of output layer, $f_o$ is the non-linear function at the output layer node, $m$ is number of nodes in the hidden layer, $w_{k,j}$ is the weight connecting $j$-th hidden node and $k$-th output node and $h_0=1$ is the bias term.
ANN is a supervised classifier and weights are determined in training phase. Generally, back propagation (BP) algorithm is used. If t-th pattern is presented by $(\mathbf{x}_t,d_t)$ where $\mathbf{x}_t$ is $t$-th input pattern and $d_t$ is the desired output or class label.
The total error in training defined as
\begin{equation}
 E(w)=\frac{1}{2}\displaystyle\sum_{j=1}^{s}(d_{t,j}-o_{t,j})^{2}
 \end{equation}

where $d_{t,j}$ and $o_{t,j}$ are the desired and actual output of $t$-th pattern at $j$-th output node, $s$ represents the nodes in the output for a given training pattern. The error is minimized by updating the
weights using the gradient descent rule:
\begin{equation}
\nabla w_{j,i}= -\eta \frac{\partial E}{\partial w_{j,i}}
\end{equation}

where $(0<\eta<1)$ is the learning rate. A small value of $\eta$ can guarantee convergence but makes learning slow. On the other hand, a large value of $\eta$ involves a rapid learning but can lead to oscillation or even divergence. To overcome this limitation, many
variations of this algorithm have been introduced for training neural networks. Other than gradient decent based BP algorithm, adaptive learning rate BP, resilient BP, Levenberg–Marquardt, and scaled conjugate gradient BP algorithms are also used for this purpose~\citep{Kandaswamy2004523}. In our work, we have used Resilient back propagation algorithm for training the neural network.

\subsection{Support vector machine (SVM)}
The SVM is a supervised machine learning technique which is based on guaranteed risk bounds of statistical learning theory known as \emph{structural risk minimization} (SRM) principle~\citep{ari2010detection} and it is used for both classification~\citep{ari2010detection} and regression~\citep{smola2004tutorial}.

Let $\{\mathbf{x}_i,y_i\}_{i=1}^{M}$ be the training examples where $\mathbf{x_i}$ is the $i$-th feature vector of dimension $d$ and ${y_i}\in{(+1,-1)}$ is the label (target output) of $\mathbf{x}_i$. The hyper plane that separates the classes is represented as,
\begin{equation}\label{eqnn}
\mathbf{w}.\mathbf{x}+b = 0,
\end{equation}

where $\mathbf{w}$ is the weight vector perpendicular to the separating hyperplane, $b$ is a scalar bias term which determines the position of hyper plane in $d$ dimensional space. Equation~\ref{eqnn} is used as inequality in the following way to separates both the classes.
\begin{equation}
\mathbf{w}.\mathbf{x}_i+b \begin{cases}
    \geq  +1 \quad \text{then}\,\, y_i=+1  \\
    \leq  -1 \quad \text{then}\,\, y_i=-1 \\
      \end{cases}
\end{equation}

\begin{figure}
\centering
\includegraphics[scale=0.25]{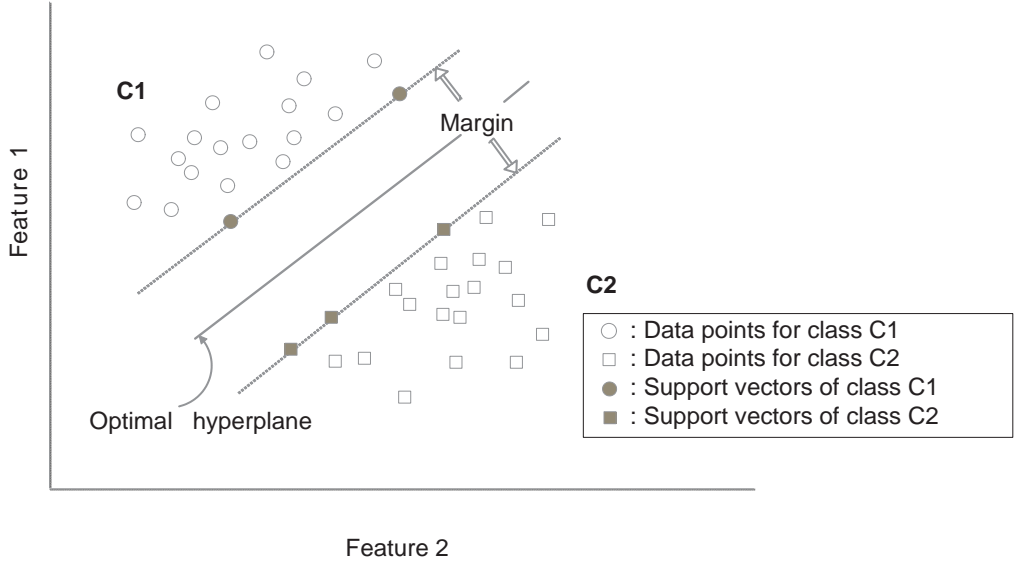}\\
\caption{An illustration of SVM classifier.}\label{SVM}
\end{figure}

Now, after including the class information, the above equations will be:
\begin{equation}
y_i(\mathbf{w}.\mathbf{x}_i+b) \geq 1 ; \quad for \; i=1,2,\ldots,M.
\end{equation}
This is for linearly separable condition. However, if a separating hyperplane does not exist in the feature space, slack variables, ${\xi_i}_{i=1}^{M}$, are introduced such that~\citep{ari2010detection},
\begin{equation}
y_i(\mathbf{w}.\mathbf{x}_i+b) \geq 1 - \xi_i; \quad \xi_i\geq 0; \quad for \; i=1,2,\ldots,M
\end{equation}

Now, to find the optimal separating hyperplane, the risk bound is minimized according to the SRM principle by the following optimization problem,
\begin{equation}\label{eqn}
\textrm{Min} \quad \frac{1}{2}\mathbf{w}.\mathbf{w} + C\sum_{i=1}^{M}\xi_i
\end{equation}

where, the parameter $C~(>0)$ is the regularization parameter that balances the importance between the maximization of the margin width and the minimization of the training error. The solution of quadratic optimization problem in Eq.~\ref{eqn}, is obtained by finding the saddle point of the Lagrange function~\citep{Ghorai2009510},
\begin{equation}
L(w,b) = \frac{1}{2}\mathbf{w}.\mathbf{w}-\displaystyle\sum_{i=0}^{M}\alpha_i{[y_i{(\mathbf{w}.\mathbf{x}_i+b)-1}]}
\end{equation}

where, $\alpha_i~(0<\alpha_i<C)$ is the Lagrange multiplier of the $i$-the data point. Note that $\alpha_i$s are non-zero only for the \emph{support vectors}, i.e., data points in the margin.

Now, the final form of the decision function that we get is as following,
\begin{equation}
f(\mathbf{x})=sgn(\mathbf{w}.\mathbf{x}+b)=sgn(\sum_{i=1}^{M}(\alpha_iy_iK(\mathbf{x}_i,\mathbf{x})+b))
\end{equation}
where, $K(.,.)$ represents the kernel function. In our work, we first use a \textit{linear inner product kernel} which is defined by,
\begin{equation}
K(\mathbf{x}_i,\mathbf{x})={\mathbf{x}_i.\mathbf{x}}
\end{equation}


In order to improve the separability of the classes, the input features are first transformed into a different domain using linear and nonlinear transformation techniques. During SVM training, maximum-margin hyperplane is obtained in the transformed feature space. In this work, along with the simple linear kernel that does not require any feature mapping, we have used \emph{Bhattacharyya}~\citep{kondor2003kernel,lee2011using} and \emph{intersection}~\citep{maji2013efficient,maji2008classification}. We select this two kernels because these are useful when the input feature is normalized histogram similar to our proposed LBP-based feature. These two kernels measure the similarity between two distributions.

\emph{Bhattacharyya kernel}: When $\mathbf{x}$ is non-negative and it represents a normalized histogram, the Bhattacharyya kernel uses \emph{Bhattacharyya coefficient} to measure the similarity, and it is defined as~\citep{kondor2003kernel},
\begin{equation}
 K(\mathbf{x}_i,\mathbf{x})=\sqrt{\mathbf{x}_i.\mathbf{x}} = (\sqrt{\mathbf{x}_i}).(\sqrt{\mathbf{x}}).
\end{equation}

\emph{Intersection kernel}: Histogram intersection kernel measures the intersection of the two normalized histograms as similarity and it is defined as~\citep{maji2013efficient},
\begin{equation}
K(\mathbf{x_i,x})=\sum_{j=1}^{d}\mathrm{min}(\mathbf{x}_i(j),\mathbf{x}(j))
\end{equation}

\section{Experimental setup}\label{Experimental setup}

\begin{table}[t]
\centering
\caption{Description of the lung sound databases used in our experiments.}
\begin{tabular}{ |l|l|l| }
\hline
Database & Sound & No. of Cycles \\ \hline
\multirow{2}{*}{Database 1} & Normal & 24  \\
 & Abnormal  & 48 (Wheeze and Crackle)  \\
\hline
\multirow{2}{*}{Database 2} & Normal & 40 \\ 
 & Abnormal & 40 (ILD) \\ 
\hline
\end{tabular}
\end{table}

\subsection{Database}\label{Label_Database}
Systematic collection of lung sound samples with the reliable ground truth is an important requirement of this research. Our database consists of recorded lung sounds obtained from three different resources: RALE database\footnote{\url{http://www.rale.ca/Repository.htm}}, Audio and Bio-signal Processing Lab (IIT Kharagpur)\footnote{\url{http://www.ecdept.iitkgp.ernet.in/index.php/home/labs/bio-sig-proc}}, Institute of Pulmocare and Research (Salt Lake, Kolkata)\footnote{\url{http://www.pulmocareindia.org/}}. The sampling frequency is $8000$ Hz in all the two cases. The recordings were done in the anterior suprasternal notch positions and trachea of
the subjects using a single channel data acquisition system described in~\citep{mondal2011reduction}. Subjects were in sitting position and at relaxing mood during recording,
and to reduce man made artifacts, stethoscope was tied with a tape on the recording site. Subjects of various age groups were involved in the recording. The signals were verified by experts. In our study, we have used two databases. The first database is categorized differentiating common three types of lung sounds. On the other hand, the sounds of other database are from subjects of ILD diseases. 
 \begin{itemize}
   \item \textbf{Database 1}: Database 1 consists of three types of lung sounds from 30 subjects - normal, crackle and wheeze. It is having 72 cycles (24 from each class) for our experiments.
   \item \textbf{Database 2}: This database includes two types of lung sounds 1) Normal and 2) ILD sound cycles. Five cycles are collected from each of eight normal subjects and eight ILD subjects.
 \end{itemize}

\subsection{Pre-processing and feature extraction}
In the pre-processing step, after reducing the effect of heart sounds~\citep{mondal2011reduction}, the lung sound cycles were extracted using a Hilbert envelope based algorithm~\citep{mondal2014detection} in the same manner as in~\citep{sengupta2016lung}. Unlike the extracted cycles  in~\citep{lozano2016automatic}, all the normal cycles were recorded from normal subjects where adventitious sounds were recorded from abnormal subjects.

Then, the sound cycles are down-sampled at $4000$ Hz because lung sounds have most of the relevant information with in 2000 Hz. Amplitude normalization is performed on cycle-by-cycle basis. For short-term feature extraction, we have used Hamming window and frame-length of 20 ms having 50\% overlap with the adjacent frames~\citep{sengupta2016lung}. MFSCs are computed using 20 mel filters, and from concatenated MFSCs, textrogram is computed using uniform LBP pattern. Note that MFCC features can be derived directly from MFSC by applying DCT.

We have also implemented four existing methods: wavelet-based~\citep{Kandaswamy2004523}, MFCC-based~\citep{sengupta2016lung}, morphological~\citep{mondal2014detection} and rational dilation wavelet based~\citep{ulukaya2016lung} technique. Wavelet coefficients are calculated using Daubechies mirror filters of order 8 (db8) with six levels of decompositions, encompassing a frequency range of $0$-$2000$ Hz. Feature vectors are computed from those coefficients of the decomposed subbands. Mean of the absolute values of the coefficients in each subband, the standard deviation of the coefficients in each subband, average power of the coefficients in each subband and ratio of computed means of adjacent subbands are calculated to formulate 27-dimensional feature vector. On the other hand, mean of conventional MFCC features, computed over all the frames of a lung sound cycle, are used to formulate a 20-dimensional MFCC-based feature. In case of morphological features, four features, i.e., kurtosis, skewness, lacunarity, and sample entropy values are computed from time domain lung sound cycles. Therefore, the feature vector dimension is four here. Now, in case of rational dilation wavelet based features, energy features that yielded best results as reported in~\citep{ulukaya2016lung} are computed from high Q-factor wavelets where dilation factor is 1.17. The dimension of the feature vector is 31.

\subsection{Classifier}
k-Nearest Neighbor (kNN): We have used MATLAB function of kNN classifier. The distance metric used here is euclidian distance. The number of neighbors is different for different features. It is chosen where the best result is obtained for a feature.

\textbf{Artificial Neural Network (ANN)}: Configuration of ANN classifier is same as in~\citep{Kandaswamy2004523,sengupta2016lung}. It has one hidden layer, and the hidden layer is consists of 40 neurons. Activation function of hidden layer and output layer is \emph{tan sigmoid} and \emph{log sigmoid}, respectively. Resilient back propagation (RP) is used for training as found efficient in lung sound classification~\citep{Kandaswamy2004523}. Average accuracy is calculated by iterating the classification method 25 times.

\textbf{Support Vector Machine (SVM)}: We have used LIBSVM library\footnote{\url{https://www.csie.ntu.edu.tw/~cjlin/libsvm/}} functions for SVM implementation. Linear, Bhattacharyya and intersection kernels are utilized in our study. Intersection kernel was found to perform better and computationally efficient in~\citep{maji2008classification}. We have used SVM penalty factor $C=1$ (related to the margin between hyper planes) in all the experiments. In case of the morphological feature, the best results were yielded for RBF kernel as reported in~\citep{mondal2014detection}. Thus, for morphological features only, we have used RBF kernel.

\subsection{Performance evaluation}
For performance evaluation, we have adopted \emph{Leave-one-out} cross validation method. In Database 1, we test one cycle at a time, and others are used as the training sample. For Database 2, we have tested cycles of one subject at a time, and cycles of other subjects are used for training. We report the classification performance using three different metrics: specificity (SPE), sensitivity (SEN) and overall accuracy (OAA). Specificity is the proportions of normal cycles that are correctly identified as normal. Sensitivity is the proportions of abnormal cycles that are correctly identified as abnormal. Finally, overall accuracy measures the number of correctly classified normal and abnormal sound cycles with respect to the total number of test samples. Those are defined below,
\begin{eqnarray}\label{normal Accuracy}
\mathrm{SPE}&=&\frac{\mathrm{No.\ of\ correctly\ classified\ normal\ cycles}}{\mathrm{Total\ no.\ of\ normal\ cycles\ under\
test}}\times 100,\nonumber\\
\end{eqnarray}
\begin{eqnarray}\label{abnormal Accuracy}
\mathrm{SEN}&=&\frac{\mathrm{No.\ of\ correctly\ classified\ abnormal\ cycles}}{\mathrm{Total\ no.\ of\ abnormal\ cycles\ under\
test}}\times 100,\nonumber\\
\end{eqnarray}
\begin{eqnarray}\label{overall Accuracy}
\mathrm{OAA}&=&\frac{\mathrm{No.\ of\ correctly\ classified\ normal\ \&\ abnormal\ cycles}}{\mathrm{Total\ no.\ of\ normal\ \&\ abnormal\ cycles\ under\
test}}\times 100.\nonumber\\
\end{eqnarray}

\section{Results and discussions}\label{result}

\subsection{Performance comparison of baseline and proposed features}
In this section, we evaluate the proposed features for lung sound classification.  We also evaluate wavelet-based~\citep{Kandaswamy2004523}, MFCC-based~\citep{sengupta2016lung}, morphology-based~\citep{mondal2014detection} and energy coefficients from rational dilation wavelet features~\citep{ulukaya2016lung}. In addition, we have evaluated MFSC-based mean features. Table~\ref{tab:acc3} describes results of lung sound classification accuracy with kNN as a classifier. Here, we use euclidean distance for similarity measure. In case of the second database, MFCC and MFSC based features perform better than LBP. But in case of Database 1, LBP performs better than other features. Lung sound classification performance on two databases using MLP-based ANN classifier are shown in Table~\ref{tab:acc1}. In most of the cases, MFCC-based mean features outperform all other methods. LBP gives poorer accuracy compared to MFCC and MFSC features. The possible reason is that this LBP histogram based features are not suitable with ANN classifier. Also, we have used the same configuration of ANN that was used in~\citep{Kandaswamy2004523}. Performance may improve if we further optimize configuration parameters of the classifier. Morphological features and energy values of rational dilation wavelet coefficients do not perform as well as MFCC, MFSC or wavelet based features in ANN classifier with the specified configuration.

\begin{table}[t]
\caption{Lung sound classification accuracy (in \%) using existing and proposed features with KNN classifier.} \label{tab:acc3}
\centerline{
\begin{footnotesize}
\begin{tabular}{|c|ccc|ccc|}
\hline
\multirow{2}{*}{Feature} & \multicolumn{3}{|c|}{Database 1} & \multicolumn{3}{|c|}{Database 2} \\
\cline{2-7}
& SPE & SEN & OAA &  SPE & SEN & OAA\\
\hline
Wavelet \citep{Kandaswamy2004523}    & 79.16 & 95.83 & 90.27 & 92.50 & 95.00 & 93.75\\
MFCC \citep{sengupta2016lung}        & 95.83 & 95.83 & 94.44 & 95.00 & 100.00 & 97.50\\
MFSC                & 95.83 & 95.83 & 94.44 &  95.00 & 100.00 & 97.50\\
Morphological features \citep{mondal2014detection}   & 87.50 & 79.17 & 77.78 & 85 & 67.50 & 76.25\\
Energy \citep{ulukaya2016lung}      & 91.67 & 95.83 & 93.05 &  82.50 & 100.00 & 91.25\\
\hline
LBP        & 100.00 & 97.91 & 98.61 & 90.00 & 100.00 & 95.00\\
\hline
\end{tabular}
\end{footnotesize}
}
\end{table}

\begin{table}[t]
\caption{Lung sound classification accuracy (in \%) using existing and proposed features with ANN classifier.} \label{tab:acc1}
\centerline{
\begin{footnotesize}
\begin{tabular}{|c|ccc|ccc|}
\hline
\multirow{2}{*}{Feature} & \multicolumn{3}{|c|}{Database 1} & \multicolumn{3}{|c|}{Database 2} \\
\cline{2-7}
& SPE & SEN & OAA &  SPE & SEN & OAA\\
\hline
Wavelet \citep{Kandaswamy2004523}    & 82.83 & 95.50 & 91.16 &  94.30 & 97.00 & 95.65\\
MFCC \citep{sengupta2016lung}        & 97.50 &97.41 & 97.22 & 92.70 & 99.70 & 96.20\\
MFSC                & 99.50 & 97.41 & 97.16 &  88.40 & 99.60 & 94.00\\
Morphological features \citep{mondal2014detection}   & 82.17 & 90.67 & 86.33 & 83.10 & 75.10 & 79.10\\
Energy \citep{ulukaya2016lung}      & 87.50 & 89.58 & 88.89 &  85.00 & 95.00 & 90.00\\
\hline
LBP        & 85.00 & 93.00 & 89.27 &  74.00 & 81.30 & 77.65\\
\hline
\end{tabular}
\end{footnotesize}
}
\end{table}

\begin{table}[t]
\caption{Lung sound classification accuracy (in \%) using existing and proposed features with SVM classifier using different kernels.} \label{tab:acc2}
\centerline{
\begin{footnotesize}
\begin{tabular}{|c|c|ccc|ccc|}
\hline
\multirow{2}{*}{Feature} &Kernel& \multicolumn{3}{|c|}{Database 1} & \multicolumn{3}{|c|}{Database 2} \\
\cline{3-8}
& Type &SPE & SEN & OAA &  SPE & SEN & OAA\\
\hline
Wavelet \citep{Kandaswamy2004523}     &Inner Product& 83.33 & 77.08 & 79.16 &  92.50 & 92.50 & 92.50\\
MFCC \citep{sengupta2016lung}         &Inner Product& 58.33  & 97.91 & 75.00 & 87.50 & 100.00 & 93.75\\
MFSC               &Inner Product& 91.67 & 97.91 & 94.44 & 87.50 & 100.00 & 93.75\\
Morphological features \citep{mondal2014detection} &RBF& 91.67 & 95.83 & 91.67 & 77.50 & 67.50 & 72.50\\
Energy \citep{ulukaya2016lung} &Inner Product& 89.67 & 97.50 & 92.33 &  87.70 & 99.80 & 93.75\\
\hline
\multirow{3}{*}{LBP}         &Inner Product& 95.83 & 97.91 & 97.22 & 85.00 & 100.00 & 92.50\\
    &Bhattacharyya& 95.83 & 97.91 & 97.22 &  90.00 & 100.00 & 95.00\\
        &Intersection& 100.00 & 97.91 & 98.61 &  85.00 & 100.00 & 92.50\\
\hline
\end{tabular}
\end{footnotesize}
}
\end{table}

Table~\ref{tab:acc2} describes results of lung sound classification accuracy with SVM as a classifier. We have used linear inner product kernel as a default system. Other than this, we have used two different kernels suitable for the proposed feature. The results shown in Table~\ref{tab:acc2} indicate that LBP-based features are more suitable with SVM classifier than the ANN-based system as shown in Table~\ref{tab:acc1}. We have found that MFSC-based mean features are better or equally good as MFCC-based mean features. The LBP-based feature with linear inner product kernel yields considerable performance in terms of sensitivity. However, its performance is not so better in terms of specificity when compared with the MFSC-based mean feature. But with Bhattacharyya kernel, its specificity is improved, particularly for Database 2. The intersection kernel also shows better specificity. But this is not consistent for both the databases. For instance, LBP-features with both Bhattacharyya and intersection kernel show better performance than popular wavelet-based features in the first database. LBP features yield better results than morphological features too in both the databases. Thus, it is found that the performance of proposed features is not only performed better than others in kNN classifier but also in SVM classifier. However, for Database 2, the wavelet-based feature is still giving the best performance in terms of specificity even with ANN as a back-end. From pathology perspective, sensitivity is more important than specificity in order to ensure that a case must not be ``missed". The proposed features are more suitable from this viewpoint.

\subsection{Further optimization of different parameters}
In the previous section, we have shown the recognition performance with an arbitrarily chosen configuration of spectral coefficients which is popularly used in speech processing context. The same was also used in our previous work on lung sound characterization~\citep{sengupta2016lung}. In this section, we further optimize the configuration parameters of the proposed feature for lung sound detection. The performance evaluations in the previous section indicate that the proposed feature outperforms others not only with kNN classifier but also with Bhattacharyya kernel based SVM classifier. We choose LBP with Bhattacharyya kernel for the rest of the analyses. Moreover, as the first database (i.e., Database 1) includes the most common sounds that may present in healthy and different diseased subjects, all optimization experiments are first conducted on this data. After that, the optimized parameters are used to evaluate the accuracies on the other database.

\begin{figure*}[t]
\centerline{\includegraphics[width=15cm,height=8cm]{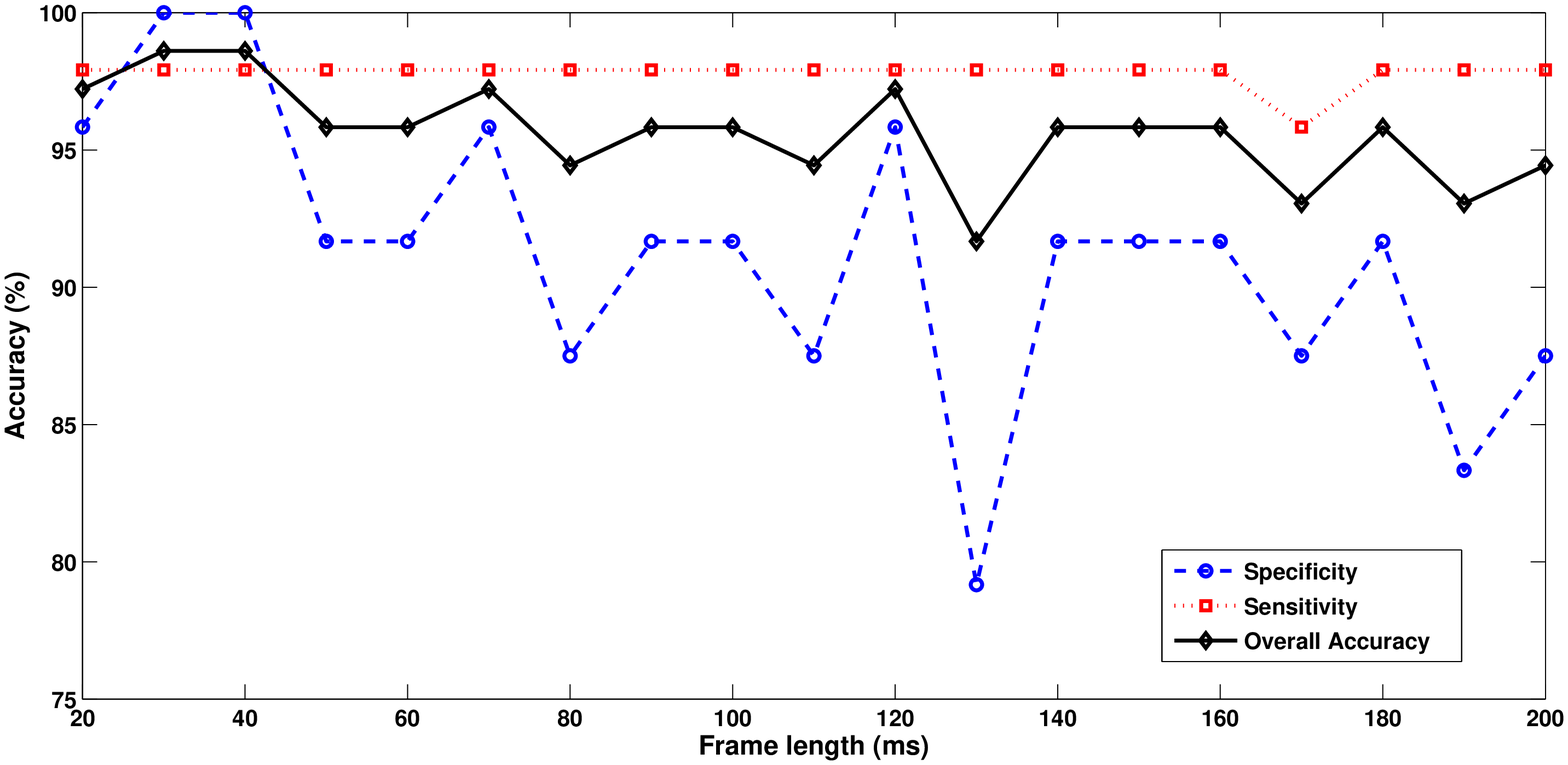}}
 \caption{Effect of frame size on lung sound recognition accuracy (\%) for proposed LBP feature with Bhattacharyya kernel and SVM classifier on Database 1.}\label{frmlngth}
\end{figure*}

\subsubsection{Optimization of frame length}
From speech signals, the short-term spectral features are generally extracted with a frame-length of 20 ms. Here, we analyze how the variations of frame length during the computations of MFSCs influence the lung sound classification accuracy. We conduct experiments by varying the frame length between 20 ms and 200 ms. The results are shown in Fig.~\ref{frmlngth}. We get best overall accuracy of 98.61\% at 30 and 40 ms frame length. As mentioned before, the default frame size of 20 ms is motivated from the speech processing application, it is not necessarily the optimum choice for lung sound analysis. We have further checked the accuracy of the proposed feature using other two kernels, we also get optimum performance at $40$ ms. Besides, from Fig.~\ref{frmlngth}, we infer that the accuracy of abnormal sound detection is less affected with the variation of frame length. On the other hand, we notice that the accuracy of normal sound detection is considerably dependent on the frame size. Interestingly, the specificity decreases with the increase of frame length. We have chosen 40 ms frame length for our next experiments with other database. In spite of change in frame-length in Database 2, the accuracies remain same for proposed feature as with 20 ms frame length (Table~\ref{tab:acc2}). Even though, the abnormal sounds are accurately detected with 100\% sensitivity, the normal sounds are not detected properly. The variation in specificity is noticed among the databases. This could be explained by several factors that affects the characteristics of normal lung sounds~\citep{reichert2008analysis}.

\subsubsection{Optimization of frame overlap}
\begin{figure*}
\centerline{\includegraphics[width=15cm,height=8cm]{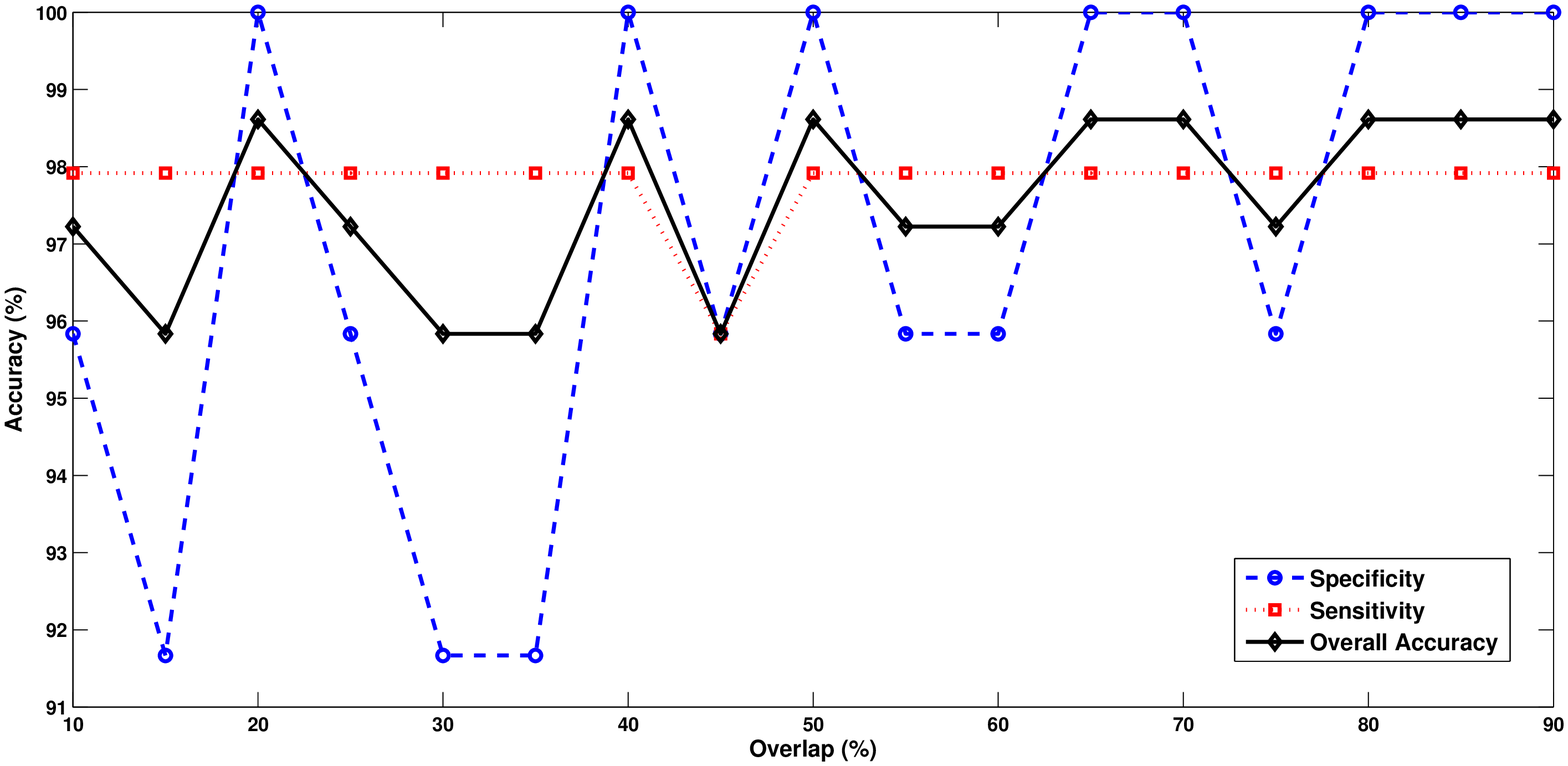}}
 \caption{Effect of percentage of frame-overlap on lung sound recognition accuracy (\%) for proposed LBP feature with Bhattacharyya kernel and SVM classifier on Database 1.}\label{ovrlp}
\end{figure*}

In the next analysis, we have varied the overlap in framing and observed its effect on the recognition performance. So far, we have used 50\% overlap. Here, we have varied it for 10-90\%. In Fig.~\ref{ovrlp}, we observe that the sensitivity is not changes over all the considered overlaps, and the variation in overall accuracy is mostly due to the detection accuracy of normal sounds. We notice that the specificity attains 100\% but it is not consistent. When percentage of overlap is set at 80\%, accuracies of both normal and abnormal sound become stable, and the best overall accuracy (i.e., 98.61\%) is achieved. Accuracies of LBP for all the three kernels saturate at percentage of overlap 90\%. Therefore, instead of taking 80\% as the optimum, we consider percentage of overlap as 90\% as the optimized value. In case of Database 2, we observe 2.5\% improvement in overall accuracy due to higher specificity in the optimized frame-overlap.

\subsubsection{Optimization of filter numbers}
\begin{figure*}
\centerline{\includegraphics[width=15cm,height=8cm]{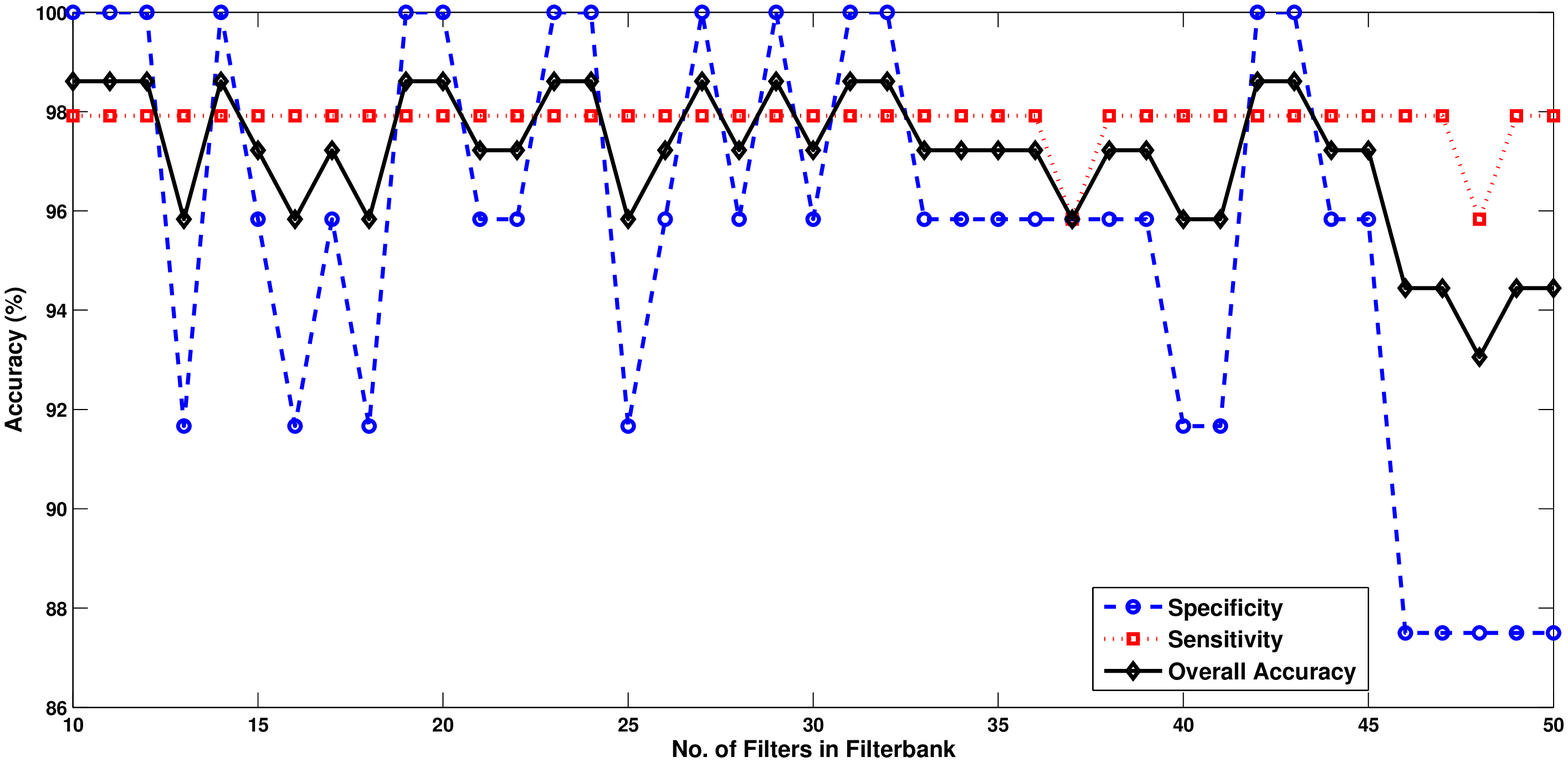}}
 \caption{Effect of number of filters in filterbank on lung sound recognition accuracy (\%) of respiratory for proposed LBP feature with Bhattacharyya kernel and SVM classifier on Database 1.}\label{fbank}
 \end{figure*}

In previous experiments, we have considered 20 filters in filterbank which is frequently used in speech processing. Here, we have varied the number of filters from 10 to 90 and observed its impact on the performance. The performances are illustrated in Fig.~\ref{fbank}. We notice that the performance is poor when the number of filters is less than 10. Surprisingly, the accuracy of abnormal sound detection, i.e., sensitivity, remains quite unchanged with the variation in number of filters. On the other hand, normal accuracy seems to be more dependent on the filter number. However, the specificity does not exhibit a consistent trend with the change of filter number. We notice that the performance of normal sound detections starts to degrade when more filters are used. Since the best accuracy is also obtained considering 20 filters (i.e., same as default value), the performance with optimum number of filters is same as the performance with default configuration as shown in Table~\ref{tab:acc2}.

\begin{figure}[t]
\centerline{\includegraphics[width=15cm,height=10cm]{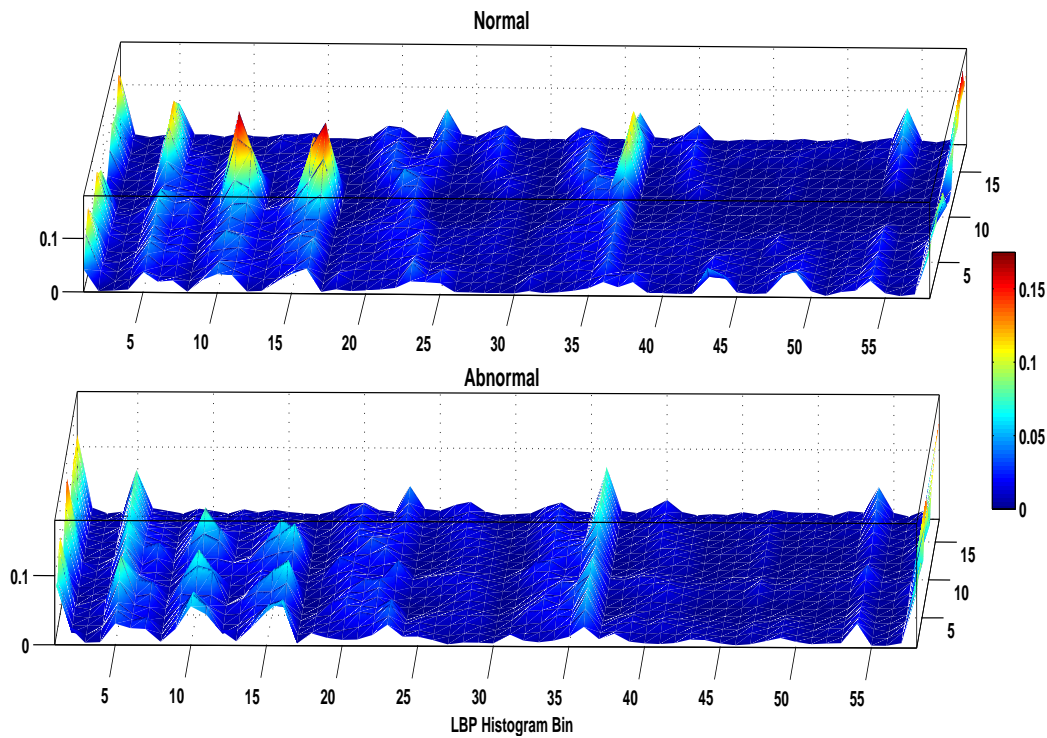}}
 \caption{Plot of average LBP features computed over all the cycles of the first database. Figure shows the features separately (i.e., without concatenation) corresponding to 18 filters.}\label{surf_plot}
\end{figure}

\begin{figure}[t]
\centerline{\includegraphics[width=16cm,height=10cm]{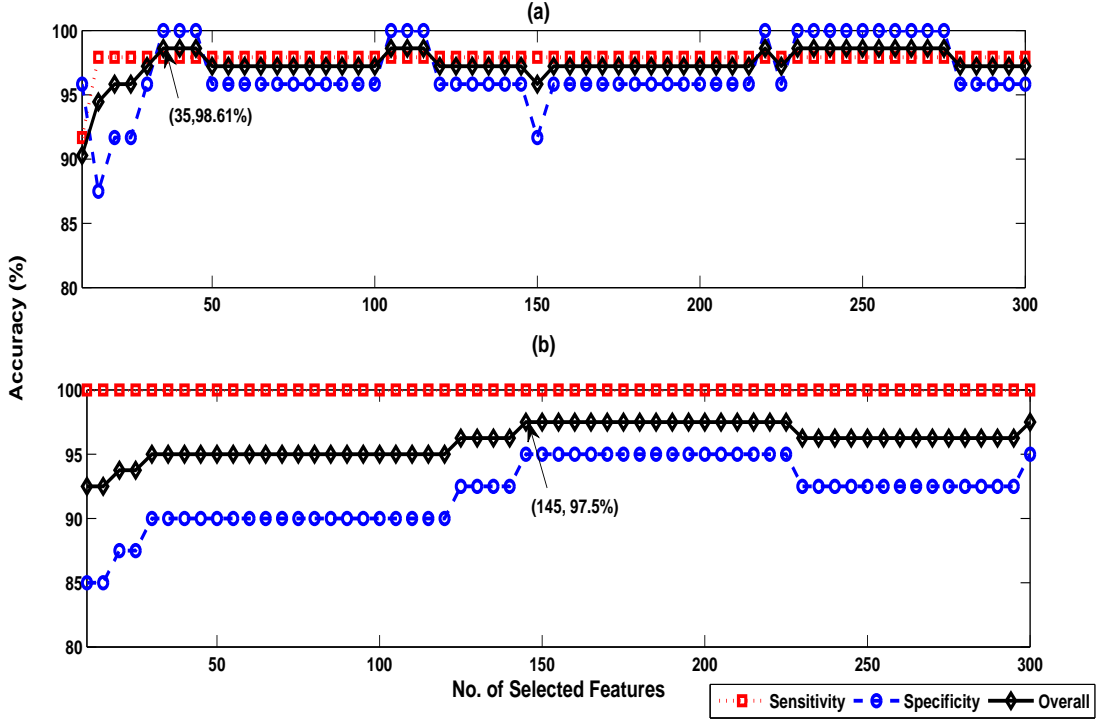}}
\caption{Lung sound recognition performance on (a) Database 1 and (b) Database 2 for different number of selected features.}\label{slctd_ftr_acc}
\end{figure}

\begin{figure}
\centerline{\includegraphics[width=14cm,height=8cm]{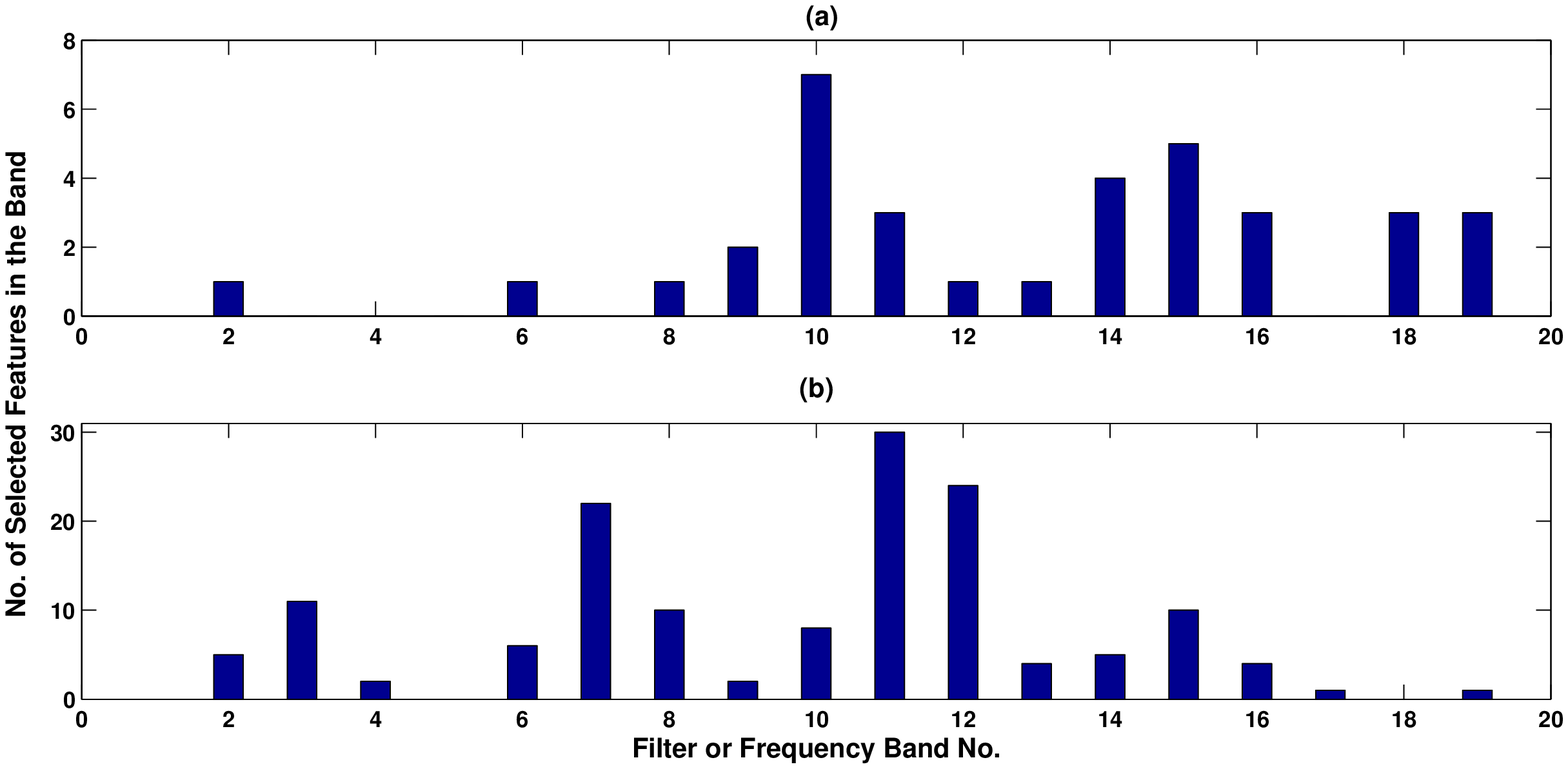}}
 \caption{Number of selected features corresponding to each filter for the discrimination of different lung sounds in two databases. The figures are shown for (a) Database 1 and (b) Database 2 where the total number of selected features are $35$, and $145$ corresponding to the maximum overall accuracy.}\label{hist_ftr_slct}
\end{figure}




%


\subsection{Feature selection}
LBP feature with Bhattacharyya and intersection kernel perform considerably well compared to standard wavelet-based features. However, its dimension is greatly expanded (1044 vs. 27). Figure~\ref{surf_plot} shows the surface plot of average LBP features which capture the energy (perceptual spectral energy) difference between neighboring frequency bands of the frame. This indicates that some features are visually more discriminative than others. In this section, we reduce the feature dimension by selecting features which are useful for discrimination. Our final objective is to identify the frequency range for effective classification. We have applied a mutual information based feature selection method known as \emph{minimum redundancy maximum relevance} (mRMR)~\citep{peng2005feature}. Then, analyzing the corresponding filters of the selected features, we find the specific frequency bands (and its temporal and spectral neighborhood) of lung sounds that have distinctive properties for different lung diseases.

Various methods use simple correlation coefficients (Fisher’s discriminant criterion, etc.) or adopt mutual information or statistical tests~\citep{guyon2003introduction} but these feature selection methods do not consider the dependency among features. Thus, for an efficient feature selection algorithm, selected features should be with minimum redundancy. The specialty of mRMR feature selection method is that it utilizes both the maximum relevance criteria, i.e.,  selecting the feature subset with the highest relevance to the target class, along with minimum redundancy criteria, i.e., reduction of redundancy among features~\citep{peng2005feature}.


Applying mRMR technique, we have selected features in the two databases, separately and conducted experiments with the selected features. For that, at first, we transform our features from continuous to categorical values~\citep{ding2005minimum,peng2005feature}. Then mRMR feature selection is employed. The features are selected on LBP histogram feature, and then the accuracies are measured using intersection kernel. Fig.~\ref{slctd_ftr_acc} shows the accuracies for two databases with respect to the number of selected features. From Fig.~\ref{slctd_ftr_acc}-(a), we observe that only 35 features give maximum overall accuracy for Database 1. It implies that the sounds in the first database is well discriminative and can be recognized with a smaller number of features.
In case of the second database, the ILD sounds are characterized well with only ten selected features which indicate that ILD sound has noticeable distinguishable characteristics as shown in~Fig.~\ref{slctd_ftr_acc}-(b). In fact, the crackle sound is a common sign presence in some diseases of ILD group which has higher discrimination ability~\citep{sengupta2016lung}. Here, the best overall accuracy is obtained by selecting 145 features. We also observe that the detection of normal sound needs more number of features in Database 2 possibly because there are different factors that affect normal sounds~\citep{reichert2008analysis}.

Finally, we have investigated the details of the selected features to understand why they are showing more distinguishable properties than the others. For this, we have computed the number of selected features corresponding to each band for MFSC, i.e., the spectral region from which eight neighborhood pixels are used for LBP analysis. Figure~\ref{hist_ftr_slct} shows the counts for each filters. Interestingly, we observe that some particular frequency bands and their neighborhood contribute more in the classification process. From Fig.~\ref{hist_ftr_slct}, we observe that even though the entire frequency range is needed for the classification of sounds in Database 1 and Database 2, some particular bands are more prominent than others.


If we look at the selected frequency bands for the first database, coefficient 10 and its neighborhood are most discriminating frequency band in case of the first database, where in case of Database 2, frequency bands 11, 12 and their neighborhoods are more distinguishable in nature. In case of Database 2, where frequency band 3 and its neighborhood play an important role, it has no role in case of Database 1. Same ways, where for Database 1, frequency band 18 and 19 have influences, no such influence is seen for Database 2. If we consider selected bands, all the bands (up 2000 Hz) is required to discriminate the features.

From the F-ratio analysis in~\citep{sengupta2016lung}, we have seen that lower frequency bands of range $<$200 Hz are contributing to distinguish different lung sounds in case of Database 1. But by using mRMR feature selection, we have observed that this portion of frequency bands are not contributing that much. The reason may be the incapability of the F-ratio analysis to remove the redundancy among features. It can be inferred that features extracted from this portion may be redundant, and thus it does not influence the lung sound detection when mRMR feature selection method is used. Only coefficient 2 and its neighborhood are enough to distinguish different lung sounds. A minimal number of features are selected that are capable of producing the best accuracy using mRMR feature selection method. Also, in F-ratio analysis, we compute the discrimination capability of a particular or individual feature. It is not necessary that individual features that have most discriminative quality individually can perform best when they perform in a group. Here, the chosen frequency bands (and their neighborhood) associated with the selected features are the features that perform best in the group and yield the best accuracy.

\section{Conclusions}\label{conclusion}
In this paper, we have studied LBP analysis of time-frequency representation of lung sounds for its classification. LBP histograms are computed for each frequency band in perceptual \emph{mel} scale and they are normalized and stacked to formulate the feature vector. The proposed feature captures the texture information of lung sounds in the time-frequency domain. We have evaluated the features with kNN, ANN and SVM classifier as back-end on two lung sound databases. Our study with different classifiers and various SVM kernels reveals that LBP-based feature with Bhattacharyya kernel and KNN classifier is superior to other methods, especially for the detection of abnormal sounds. The configuration parameters of proposed features are further optimized experimentally, and their performance is measured. During this process, it is observed that the optimum  analysis window length for short-term feature extraction of lung sound is moderately longer than the frequently used window length in speech analysis. Besides, performance is better when the overlap is more. We also have noticed, in most of the cases, detection of normal sound is more crucial, and specificity is quite sensitive to the change of configuration parameters. It is found that detailed information is needed to detect normal sound more accurately. To reduce the computational load, we have used a well-known feature selection technique which reduces feature vector dimension without degradation in the results. Finally, it is seen that while vital information is captured under 1000 Hz, some frequency bands of more than 1000 Hz also influence the results.\par
The current study is conducted with databases of limited size due to the unavailability of a large amount of data with reliable ground-truths. The newly investigated method should be validated with a larger database. The proposed feature captures relative information of the subbands whereas their absolute values are apparently not in use. Fusion methods can be explored for combining different kinds of information using system fusion techniques.

\section*{Conflict of interest statement}
The authors declare that they have no conflict of interest.

\section*{Acknowledgement}
The work was supported by Ministry of Human Resource Development, Government of India. We are thankful to Dr. Ashok Mondal for his help in the database preparation. We would also like to thank the reviewers for their careful reading of the paper and helpful comments.

\section*{References}
\bibliographystyle{elsarticle-num}

\bibliography{latexbib}

\end{document}